\documentclass[preprint,aps,12pt,amsmath,amssymb,showkeys,nofootinbib,tightenlines,superscriptaddress]{revtex4-1}

\usepackage{amssymb}
\usepackage{amsfonts}
\usepackage{latexsym}
\usepackage{pdfsync}
\usepackage[T1]{fontenc}
\usepackage{epsf,amsfonts}
\usepackage{epsfig}
\usepackage{amsthm}
\usepackage{amsmath}
\usepackage{graphicx}
\usepackage{subfigure}
\usepackage{color}
\usepackage{slashed}
\usepackage{mathrsfs}
\usepackage{float}
\usepackage{color}
\usepackage{esvect}
\usepackage{mathtools}
\usepackage{bbm}

\newcommand{\VEV}[1]{\left\langle #1\right\rangle}

\newcommand{\tr}{\mathop{\rm tr}\nolimits}

\newcommand{\MeV}{\;\text{MeV}}
\newcommand{\GeV}{\;\text{GeV}}

\begin{document}

\title{Equation of state and chiral transition in soft-wall AdS/QCD with more realistic gravitational background}

\author{Zhen Fang}
\email{zhenfang@hnu.edu.cn}
\affiliation{Department of Applied Physics, School of Physics and Electronics, Hunan University, Changsha 410082, China}

\author{Yue-Liang Wu}
\email{ylwu@itp.ac.cn}
\affiliation{CAS Key Laboratory of Theoretical Physics, Institute of Theoretical Physics, Chinese Academy of Sciences, Beijing 100190, China}
\affiliation{International Centre for Theoretical Physics Asia-Pacific (ICTP-AP), University of Chinese Academy of Sciences, Beijing 100049, China}

\date{\today}

\begin{abstract}
We construct an improved soft-wall AdS/QCD model with a cubic coupling term of the dilaton and the bulk scalar field. The background fields in this model are solved from the Einstein-dilaton system with a nontrivial dilaton potential, which has been shown to be able to reproduce the equation of state from lattice QCD with two flavors. The chiral transition behaviors are investigated in the improved soft-wall AdS/QCD model with the solved gravitational background, and the crossover transition can be realized. Our study provides a possibility to address the deconfining and chiral phase transitions simultaneously in the bottom-up holographic framework. 
\end{abstract}

\maketitle

\section{Introduction}\label{introduce1}

As is well known, quantum chromodynamics (QCD) describes the strong interaction of quarks and gluons. Due to asymptotic freedom, the method of perturbative quantum field theory can be used to study high-energy properties of QCD matters in the ultra-violet (UV) region. However, the strong coupling nature of QCD at low energies makes the perturbative method invalid to tackle the nonperturbative problems of QCD. Quark confinement and chiral symmetry breaking are two essential features of low-energy QCD, and the related physics have attracted a great deal of interests since many years ago. QCD phase transition is such a field that has opened a window for us to look into the low-energy physics of QCD \cite{Stephanov:2007fk}. As the temperature increases, we know that the QCD matters undergo a crossover transition from the hadronic state to the state of quark-gluon plasma (QGP), along with the deconfining process of the partonic degrees of freedom and the restoration of chiral symmetry \cite{Aoki:2006we,Bazavov:2011nk,Bhattacharya:2014ara}.

Many nonperturbative methods have been developed to study the QCD phase transition and the issues of low-energy hadron physics \cite{Roberts:2000aa,Braun:2006jd,Fischer:2009wc,Fukushima:2003fw}. As a powerful method, lattice QCD is widely used to tackle the low-energy QCD problems from the first principle. However, there are limitations on this method, such as in the case of nonzero chemical potential, because of the sign problem. In recent decades, the anti-de Sitter/conformal field theory (AdS/CFT) correspondence has provided a powerful tool for us to study the low-energy physics of QCD by the holographic duality between a weakly coupled gravity theory in asymptotic AdS$_5$ spacetime and a strongly coupled gauge field theory on the boundary \cite{Maldacena:1997re,Gubser:1998bc,Witten:1998qj}. Large amounts of researches have been done in this field, following either the top-down approach or the bottom-up approach \cite{Kruczenski:2003uq,Sakai:2004cn,Sakai:2005yt,DaRold:2005mxj,Erlich:2005qh,deTeramond:2005su,Karch:2006pv,Csaki:2006ji,Cherman:2008eh,Gherghetta:2009ac,Kelley:2010mu,Sui:2009xe,Sui:2010ay,Fujita:2009wc,Fujita:2009ca,Colangelo:2009ra,Colangelo:2011sr,Cui:2013xva,Cui:2014oba,Li:2012ay,Li:2013oda,Shuryak:2004cy,Brodsky:2014yha,Tannenbaum:2006ch,Policastro:2001yc,Cai:2009zv,Cai:2008ph,Sin:2004yx,Shuryak:2005ia,Nastase:2005rp,Nakamura:2006ih,Sin:2006pv,Janik:2005zt, Herzog:2006gh,Gursoy:2007cb,Gursoy:2007er,Gursoy:2008bu,Gursoy:2008za,Li:2014hja,Li:2014dsa,Fang:2016uer,Fang:2016dqm,Fang:2016cnt,Evans:2016jzo,Mamo:2016xco,Dudal:2016joz,Dudal:2018rki,Ballon-Bayona:2017dvv,Chen:2018msc,Attems:2016ugt,Attems:2017zam}. 

Holographic studies in the top-down approach have shown that the simplest nonsupersymmetric deformation of AdS/CFT with nontrivial dilaton profiles can reproduce the confining properties of QCD \cite{Gubser:1999pk,Brevik:2005fs}, and also realize the pattern of chiral symmetry breaking with quarks mimicked by the D7-brane probes \cite{Karch:2002sh,Kruczenski:2003be,Erdmenger:2007cm,Babington:2003vm,Ghoroku:2004sp}. However, it is not clear how to generate crossover transition indicated from lattice QCD in the top-down framework. Actually, AdS/CFT per se is inadequate to give a complete description for thermodynamics of QCD because of its semi-classical character inherited from the type IIB string theory in the low-energy approximation and the large $N$ limit. The string-loop corrections have to be considered in order to give an adequate account for thermal QCD. Nevertheless, the qualitative description by such a holographic approach is still meaningful and indeed has provided many insights in our study of low-energy QCD.

It has been shown that the deconfinement in the pure gauge theory corresponds to a Hawking-Page phase transition between a thermal AdS space and a black hole configuration \cite{Hawking:1982dh,Witten:1998zw,Herzog:2006ra}. However, many works from bottom-up approach tell us that we can use a bulk gravity system with a nontrivial dilaton profile to characterize the equation of state and the deconfining behaviors of QCD \cite{Gubser:2008yx,Gubser:2008ny,Gubser:2008sz,Andreev:2009zk,Noronha:2009ud,Finazzo:2013efa,Finazzo:2014zga,Yaresko:2013tia,Yaresko:2015ysa,Colangelo:2010pe,Li:2011hp,He:2013qq,Yang:2014bqa,Fang:2015ytf,Rougemont:2017tlu,Li:2017ple,Zollner:2018uep,ChenXun:2019zjc}. Moreover, unlike the holographic studies of pure gauge theory, the crossover transition in these bottom-up models is only related to the black hole solution solved from the Einstein-dilaton(-Maxwell) system, which is contrary to the usual claim that the black hole is dual to the deconfined phase at high temperatures. As we cannot expect to make use of two distinct bulk geometries to generate a smooth crossover transition in AdS/QCD \cite{Colangelo:2009ra}, it seems more natural to give a description on thermal QCD properties only in terms of the black hole solution. However, in this case we must make sure that the black hole is stable when compared with the thermal gas phase.

In the bottom-up approach, the soft-wall AdS/QCD model provides a concise framework to address the issues on chiral transition \cite{Karch:2006pv}. However, it has been shown that the original soft-wall model lacks spontaneous chiral symmetry breaking \cite{Karch:2006pv,Colangelo:2011sr}. The chiral transitions in the two-flavor case have been studied in a modified soft-wall AdS/QCD model, where the second-order chiral phase transition in the chiral limit and the crossover transition with finite quark masses are first realized in the holographic framework \cite{Chelabi:2015gpc,Chelabi:2015cwn}. In Ref. \cite{Fang:2016nfj}, we proposed an improved soft-wall model which can generate both the correct chiral transition behaviors and the light meson spectra in a consistent way. The generalizations to the $2+1$ flavor case have been considered in Ref. \cite{Li:2016smq,Bartz:2017jku,Fang:2018vkp}, and the quark-mass phase diagram that is consistent with the standard scenario can be reproduced. The case of finite chemical potential has also been investigated \cite{Bartz:2016ufc,Fang:2018axm,Fang:2019lmd}, and the chiral phase diagram containing a critical end point can be obtained from the improved soft-wall AdS/QCD model with $2+1$ flavors \cite{Fang:2018axm,Fang:2019lmd}. 

It should be noted that the AdS-Schwarzschild black-hole background has been used in most of studies of chiral transition at zero chemical potentials. However, such an AdS black-hole solution is dual to a conformal gauge theory, which cannot generate the QCD equation of state without breaking the conformal invariance \cite{Gubser:2008yx}. As just mentioned, one can resort to the Einstein-dilaton system with a nontrivial dilaton profile to rescue this issue. So we wonder whether the correct chiral transition behaviors can still be obtained from a soft-wall AdS/QCD model with a solved gravitational background from the Einstein-dilaton system. In this work, we shall consider this issue and try to combine the description of chiral transition with that of the equation of state which signifies deconfinement in a unified holographic framework.

This paper is organized as follows. In Sec. \ref{sec-eos}, we consider an Einstein-dilaton system with a nontrivial dilaton potential, which can produce the equation of state consistent with lattice results in the two-flavor case. The vacuum expectation value (VEV) of the Polyakov loop will also be computed in such a background system, and will be compared with the lattice data. In Sec. \ref{sec-SW}, we propose an improved soft-wall AdS/QCD model with a cubic coupling term of the dilaton and the bulk scalar field, and the chiral transition behaviors will be considered in the two-flavor case. It will be seen that the crossover behaviors of chiral transition can be realized in this model. The parameter dependence of chiral transition will also be investigated. In Sec. \ref{sec-conc}, we give a brief summary of our work and conclude with some remarks.

\section{QCD equation of state from holography}\label{sec-eos}

\subsection{The Einstein-dilaton system}\label{gravi-dilat}

In the previous works, we proposed an improved soft-wall AdS/QCD model with a running bulk scalar mass $m_5^2(z)$, which gives quite a good characterization for the chiral transition in both the two-flavor and the $2+1$ flavor case \cite{Fang:2016nfj,Fang:2018vkp}. However, the AdS-Schwarzschild black hole presumed in this model cannot describe the thermodynamical behaviors of QCD equation of state and other equilibrium quantities which show obvious violation of conformal invariance \cite{Gubser:2008yx}. In order to acquire these basic features of thermal QCD, we need to construct a proper gravity background other than the AdS-type black hole to break the conformal invariance of the dual gauge theory. The minimal action of such a background system is given in the string frame as
\begin{align}\label{gravity-act-str1}
S_g &= \frac{1}{2\kappa_5^2}\int d^5x\sqrt{-g}e^{-2 \phi }\left[R +4(\partial\phi)^2 -V(\phi)\right] ,
\end{align}
where $\kappa_5^2=8\pi G_5$, and a dilaton field $\phi$ has been introduced to produce relevant deformations of the dual conformal field theory. The dilaton $\phi(z)$ is assumed to depend only on the radial coordinate $z$. The key point of this model is to find a particular form of the dilaton potential $V(\phi)$ with necessary ingredients to describe the QCD thermodynamics, such as the equation of state.

The metric of the bulk geometry in the string frame can be written as
\begin{align}\label{stringmetric}
ds^2 &= \frac{L^2 e^{2 A_S(z)}}{z^2} \left(-f(z)dt^2 + dx^i dx^i +\frac{dz^2}{f(z)}\right)
\end{align}
with the asymptotic structure of AdS$_5$ spacetime at $z\to0$ to guarantee the UV conformal behavior of the dual gauge theory on the boundary. We take the $\rm{AdS}$ radius $L=1$ for convenience. To simplify the calculation, we will work in the Einstein frame with the metric ansatz
\begin{align}\label{einst-metric}
ds^2 &= \frac{L^2 e^{2 A_E(z)}}{z^2} \left(-f(z)dt^2 + dx^i dx^i +\frac{dz^2}{f(z)}\right) .
\end{align}
The warp factors in the two frames are related by $A_S=A_E+\frac{2}{3}\phi$, in terms of which the background action in the Einstein frame can be obtained from the string-frame action (\ref{gravity-act-str1}) as 
\begin{align}\label{gravity-act-ein1}
S_g &=\frac{1}{2\kappa_5^2}\int d^5x\sqrt{-g_E}\left[R_E -\frac{4}{3}(\partial\phi)^2 -V_E(\phi)\right]
\end{align}
with $V_E(\phi) \equiv e^{\frac{4\phi}{3}}V(\phi)$ (the subscript $E$ denotes the Einstein frame).

\subsection{The EOM with a nontrivial dilaton potential}

The independent Einstein equations can be derived by the variation of the action (\ref{gravity-act-ein1}) with respect to the metric $g_{MN}$,
\begin{align}
& f'' +3A_E'f' -\frac{3}{z}f' =0 ,  \label{fz-eom2}    \\
& A_E'' +\frac{2}{z}A_E' -A_E'^2 +\frac{4}{9}\phi'^2 =0 .  \label{AE-eom2}
\end{align}
The equation of motion (EOM) of the dilaton $\phi$ in the Einstein frame can also be derived as
\begin{align} \label{dilaton-eom2}
\phi'' +\left(3 A_E' +\frac{f'}{f}-\frac{3}{z}\right)\phi' -\frac{3e^{2A_E}\partial_{\phi}V_E(\phi)}{8z^2f} =0 .
\end{align}
Given the dilaton potential $V_E(\phi)$, the numerical solution of the background fields $A_E$, $f$ and $\phi$ can be solved from the coupled differential equations (\ref{fz-eom2}), (\ref{AE-eom2}) and (\ref{dilaton-eom2}). 

Although there are few constraints on the form of the dilaton potential from the top-down approach of AdS/QCD, it has been shown that a proper $V_E(\phi)$ can be constructed from bottom up to describe the equation of state of the strongly coupled QGP \cite{Gubser:2008yx,Gubser:2008ny}. Near the boundary, the bulk geometry should approach the AdS$_5$ spacetime that corresponds to a UV fixed point of the dual gauge theory. This requires that the dilaton potential at UV has the following asymptotic form:
\begin{align}\label{ein-V-phi-UV}
V_c(\phi_c\to0) \simeq -\frac{12}{L^2}+\frac{1}{2}m^2\phi_c^2+\mathcal{O}(\phi_c^4)
\end{align}
with the rescaled dilaton field defined by $\phi_c =\sqrt{\frac{8}{3}}\phi$, in terms of which the action (\ref{gravity-act-ein1}) can be recast into the canonical form
\begin{align}\label{cano-gra-act1}
S_g &=\frac{1}{2\kappa_5^2}\int d^5x\sqrt{-g_E}\left[R_E -\frac{1}{2}(\partial\phi_c)^2 -V_c(\phi_c)\right]
\end{align}
with $V_c(\phi_c) =V_E(\phi)$. As argued in Ref. \cite{Gubser:2008ny}, the dilaton potential at IR takes an exponential form $V_c(\phi_c) \sim V_0e^{\gamma\phi_c}$ with $V_0<0$ and $\gamma>0$ in order to yield the Chamblin-Reall solution, whose adiabatic generalization links the QCD equation of state to the specific form of $V_c(\phi_c)$.

According to AdS/CFT, the mass squared of $\phi_c$ is related to the scaling dimension $\Delta$ of the dual operator on the boundary by $m^2 L^2=\Delta(\Delta-4)$ \cite{Erlich:2005qh}. We only consider the case of $2<\Delta<4$, which corresponds to the relevant deformations satisfying the Breitenlohner-Freedman (BF) bound \cite{Gubser:2008yx,Noronha:2009ud}. It is usually assumed that the dilaton field $\phi_c$ is dual to the gauge-invariant dimension-4 glueball operator $\tr F^2_{\mu\nu}$, yet other possibilities such as a dimension-2 gluon mass operator have also been considered \cite{Li:2013oda}. Following Ref. \cite{Gubser:2008yx}, we try to match QCD at some intermediate semi-hard scale, where the scaling dimension of $\tr F^2_{\mu\nu}$ would have a smaller value than $4$. One remark is that the asymptotic freedom cannot be captured in this way, but will be replaced by conformal invariance when above the semi-hard scale \cite{Gubser:2008yx}. The full consideration might go beyond the supergravity approximation. In this work, we just take $\Delta=3$, which has been shown to be able to describe the equation of state from lattice QCD with $2+1$ flavors \cite{Finazzo:2013efa,Finazzo:2014zga}, and which is also easier to implement in the numerical calculation. We note that other values of $\Delta$ can also be used to mimick the QCD equation of state, and this is by and large determined by the particular form of the dilaton potential and the specific parameter values \cite{Gubser:2008yx,Noronha:2009ud}. The main aim of our work is to investigate the chiral properties based on a gravitational background which can reproduce the QCD equation of state. Thus we will not delve into the possible influence of the value of $\Delta$ on the results considered in our work. Following the studies in Ref. \cite{Gubser:2008yx}, we choose a relatively simple dilaton potential which satisfies the required UV and IR asymptotics,
\begin{align}\label{dilaton-poten1}
V_c(\phi_c) =\frac{1}{L^2}\left(-12\cosh\gamma\phi_c +b_2\phi_c^2 +b_4\phi_c^4\right) ,
\end{align}
where $\gamma$ and $b_2$ are constrained by the UV asymptotic form (\ref{ein-V-phi-UV}) as
\begin{align}\label{gamma-b2}
b_2 =6\gamma^2 +\frac{\Delta(\Delta-4)}{2} =6\gamma^2 -\frac{3}{2} .
\end{align}
The dilaton potential $V_E(\phi)$ has the form
\begin{align}\label{VE-VEc}
V_E(\phi) =V_c(\phi_c) =V_c(\sqrt{8/3}\,\phi) .
\end{align}

We will see that the Einstein-dilaton system given above can also be used to mimick the two-flavor lattice results of the QCD equation of state, whereby the dilaton potential $V_E(\phi)$ and the background geometry can be reconstructed for the two-flavor case.

\subsection{Equation of state}\label{sec-EOS}

Now we come to the equation of state in the Einstein-dilaton system with the given form of dilaton potential (\ref{dilaton-poten1}). First note that the background geometry has an event horizon at $z=z_h$ which is determined by $f(z_h)=0$. In terms of the metric ansatz (\ref{einst-metric}), the Hawking temperature $T$ of the black hole is given by
\begin{align}\label{Hawking-T}
T=\frac{|f'(z_h)|}{4\pi},
\end{align}
and the entropy density $s$ is related to the area of the horizon,
\begin{align} \label{entropy}
 s =\frac{e^{3A_E(z_h)}}{4 G_5 z_h^3}.
\end{align}
Thus we can compute the speed of sound $c_s$ by the formula
\begin{align} \label{soundspeed}
c_s^2=\frac{d\,\mathrm{log}T}{d\,\mathrm{log}s}.
\end{align}
Moreover, the pressure $p$ can be obtained from the thermodynamic relation $s=\frac{\partial p}{\partial T}$ as
\begin{align} \label{p-T}
p =-\int_{\infty}^{z_h} s(\bar{z}_h)T'(\bar{z}_h) d\bar{z}_h.
\end{align}
The energy density $\varepsilon=-p+sT$ and the trace anomaly $\varepsilon-3p$ can also be computed. Then we can study the temperature dependence of the equation of state in such an Einstein-dilaton system. As we constrain ourselves to the two-flavor case, the equation of state from lattice QCD with two flavors will be used to construct the dilaton potential $V_E(\phi)$ \cite{Burger:2014xga}.

Instead of implementing the numerical procedure elucidated in Ref. \cite{Gubser:2008ny}, we will directly solve the background fields from Eqs. (\ref{fz-eom2}), (\ref{AE-eom2}) and (\ref{dilaton-eom2}). To simplify the computation, note that Eq. (\ref{fz-eom2}) can be integrated into a first-order differential equation
\begin{align}\label{fz-eom3}
f' +f_c\,e^{-3A_E}z^3 = 0 ,
\end{align}
where $f_c$ is the integral constant. In view of $\Delta=3$, the UV asymptotic forms of the background fields at $z\to0$ can be obtained as
\begin{align}\label{BG-uv-asymp}
& A_E(z) = -\frac{2p_1^2}{27} z^2 +\cdots ,       \nonumber\\
& f(z) = 1 -\frac{f_c}{4} z^4 +\cdots ,      \\
& \phi(z) = p_1 z+p_3 z^3 +\frac{4p_1^3}{9}\left(12b_4 -6\gamma^4 +1\right)z^3\log z +\cdots        \nonumber
\end{align}
with three independent parameters $p_1, p_3$ and $f_c$. As we have $f(z_h)=0$, to guarantee the regular behavior of $\phi(z)$ near the horizon, Eq. (\ref{dilaton-eom2}) must satisfy a natural IR boundary condition at $z=z_h$,
\begin{align} \label{phi-eom-BC}
\left[f'\phi' -\frac{3e^{2A_E}}{8z^2}\partial_{\phi}V_E(\phi)\right]_{z=z_h} =0 .
\end{align}

\begin{figure*}
\begin{center}
\includegraphics[width=68mm,clip=true,keepaspectratio=true]{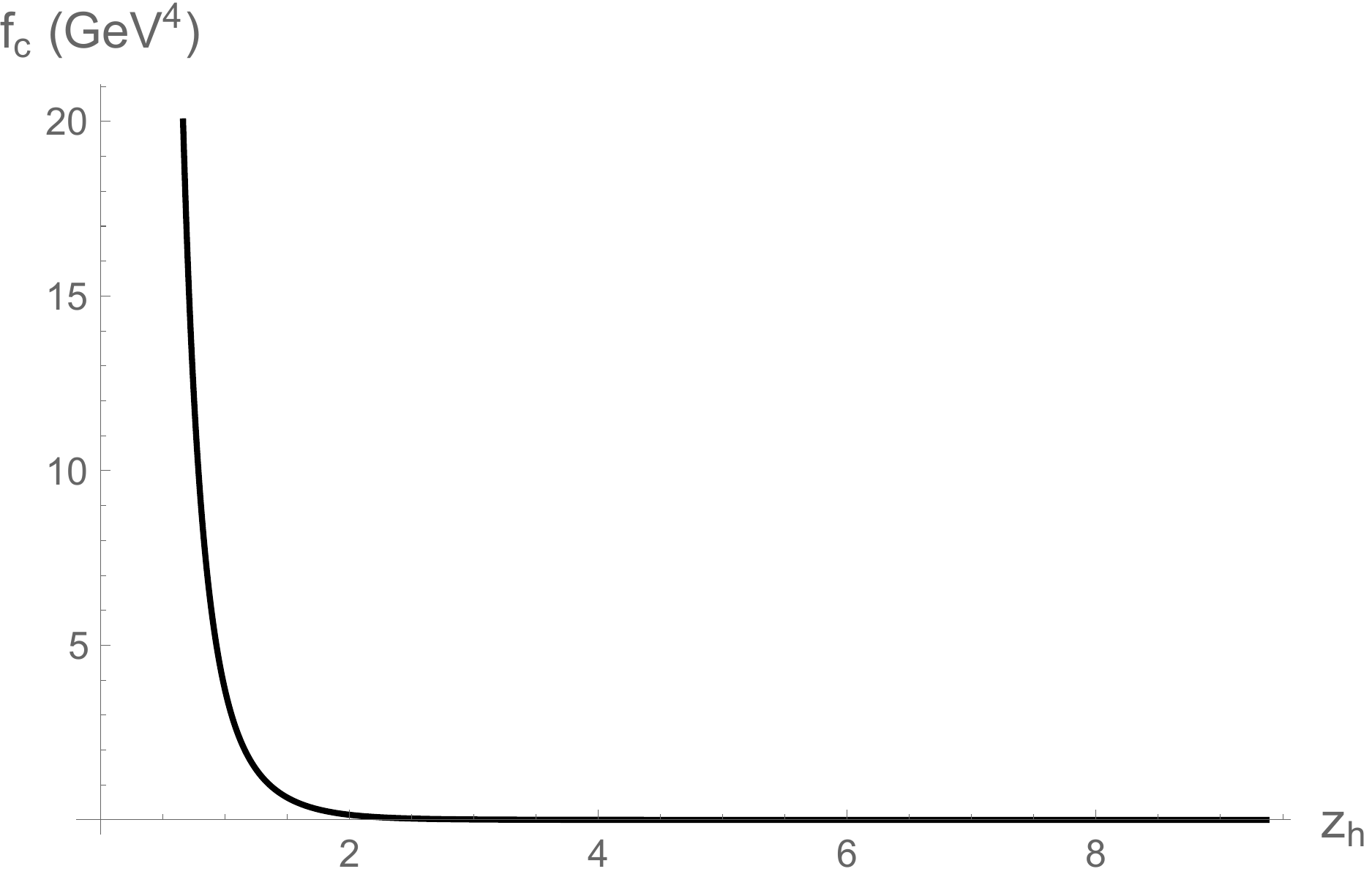}
\hspace*{0.6cm}
\includegraphics[width=68mm,clip=true,keepaspectratio=true]{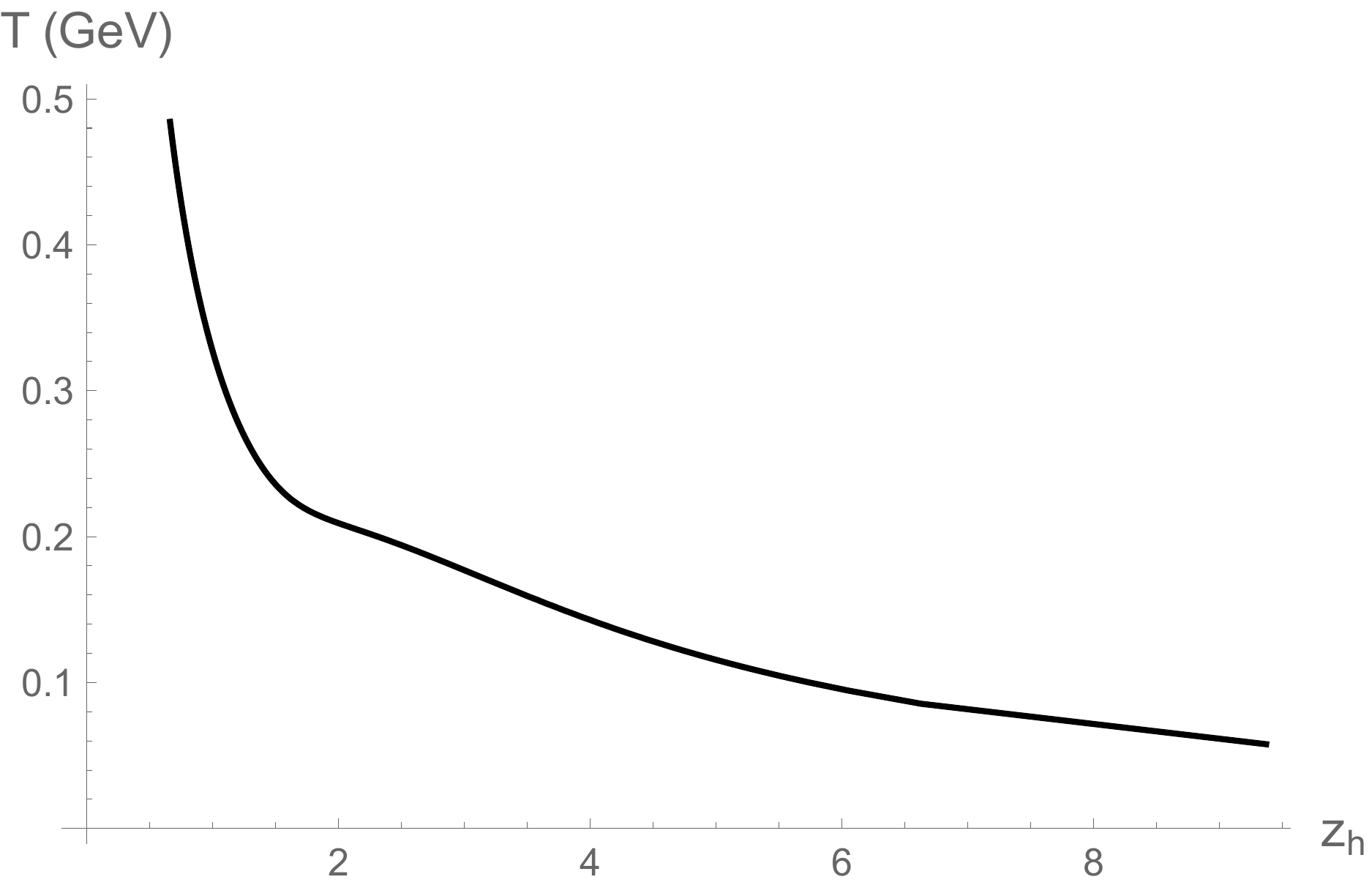}
\vskip -0.5cm \hskip 0.7 cm
\end{center}
\caption{The $z_h$-dependence of the parameter $f_c$ and the temperature $T$.}
\label{fc-T-zh}
\end{figure*}

\begin{figure}
\centering
\includegraphics[width=75mm,clip=true,keepaspectratio=true]{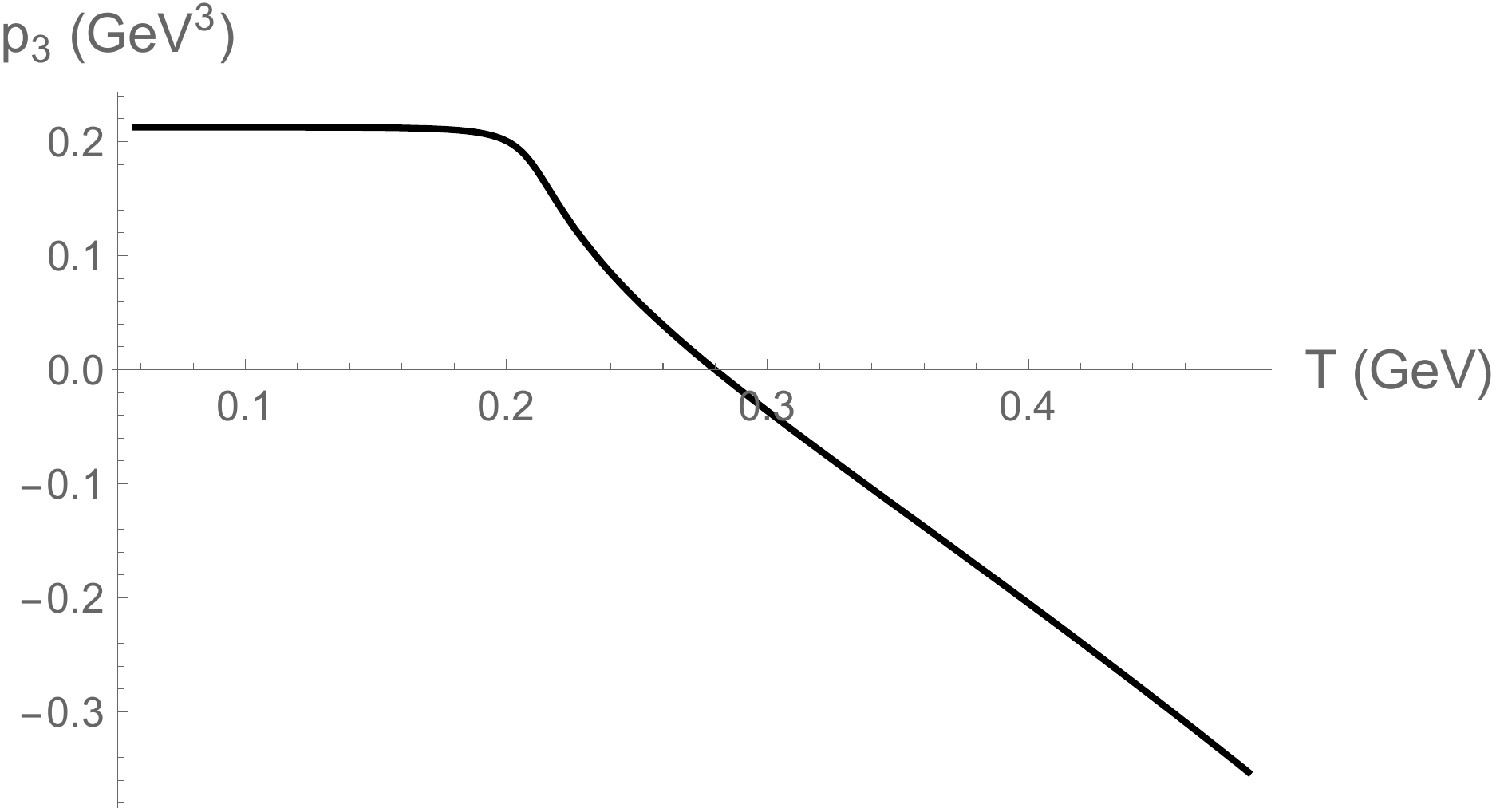}
\vskip 0.3cm
\caption{The temperature dependence of the parameter $p_3$.} 
\label{fig-p3-T}
\end{figure}

With the UV asymptotic form (\ref{BG-uv-asymp}) and the IR boundary condition (\ref{phi-eom-BC}), the background fields $f$, $A_E$ and $\phi$ can be solved numerically from Eqs. (\ref{AE-eom2}), (\ref{dilaton-eom2}) and (\ref{fz-eom3}). We find that the dilaton potential (\ref{dilaton-poten1}) with $\gamma=0.55$, $b_2=0.315$ and $b_4=-0.125$ can well reproduce the two-flavor lattice QCD results of the equation of state. Note that $\gamma$ and $b_2$ are related by the formula (\ref{gamma-b2}). The parameter $p_1=0.675 \GeV$ is also fitted by the lattice results, and the 5D Newton constant is just taken as $G_5=1$ in our consideration. The parameters $p_3$ and $f_c$ in (\ref{BG-uv-asymp}) are constrained by the IR boundary condition (\ref{phi-eom-BC}), and thus depend on the horizon $z_h$ or the temperature $T$. We show the $z_h$-dependence of the parameter $f_c$ and the temperature $T$ in Fig. \ref{fc-T-zh}. Both $f_c$ and $T$ decrease monotonically towards zero with the increase of $z_h$, which implies that our black hole solution persists in all the range of $T$. We also show the parameter $p_3$ as a function of $T$ in Fig. \ref{fig-p3-T}, where we can see that $p_3$ varies very slowly in the range of $T= 0\sim 0.2 \GeV$, then decreases and goes to negative values at around $T\simeq 0.28 \GeV$.

\begin{figure*}
\begin{center}
\includegraphics[width=68mm,clip=true,keepaspectratio=true]{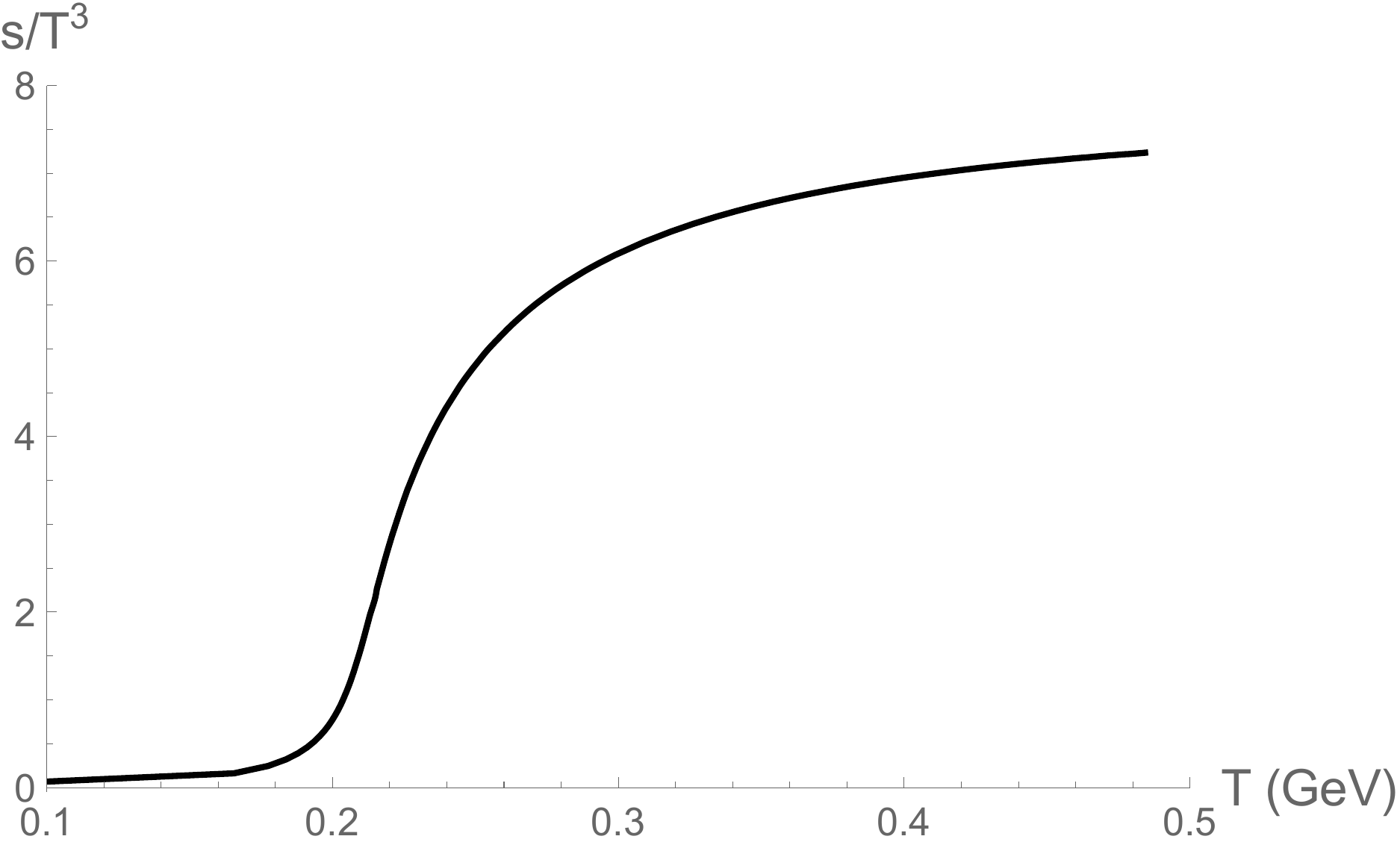}
\hspace*{0.6cm}
\includegraphics[width=68mm,clip=true,keepaspectratio=true]{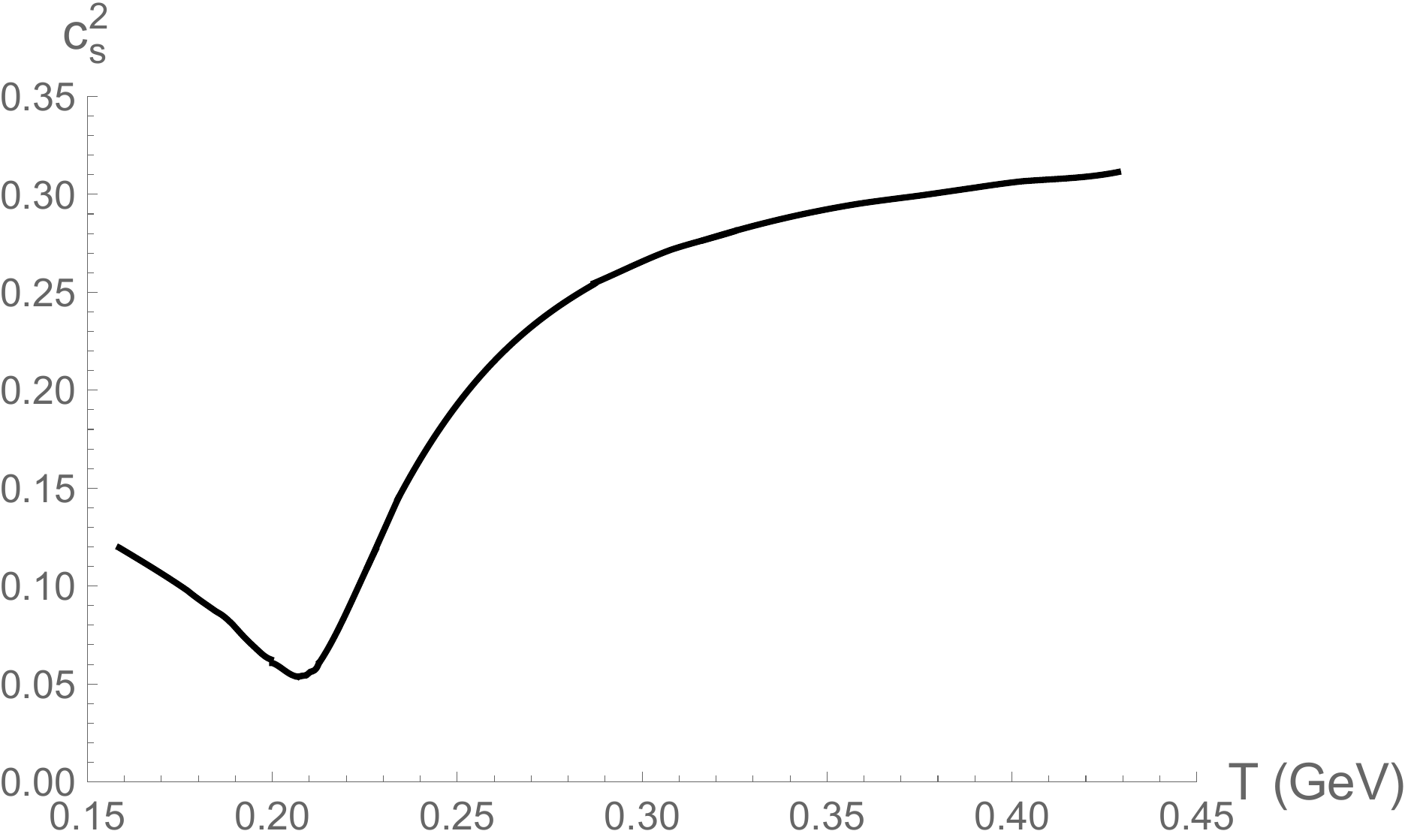}
\vskip -0.5cm \hskip 0.7 cm
\end{center}
\caption{The model results of the entropy density $s/T^3$ (left panel) and the speed of sound squared $c_s^2$ (right panel) as functions of $T$ obtained from the Einstein-dilaton system.}
\label{s-cs-T}
\end{figure*}

\begin{figure*}
\begin{center}
\includegraphics[width=68mm,clip=true,keepaspectratio=true]{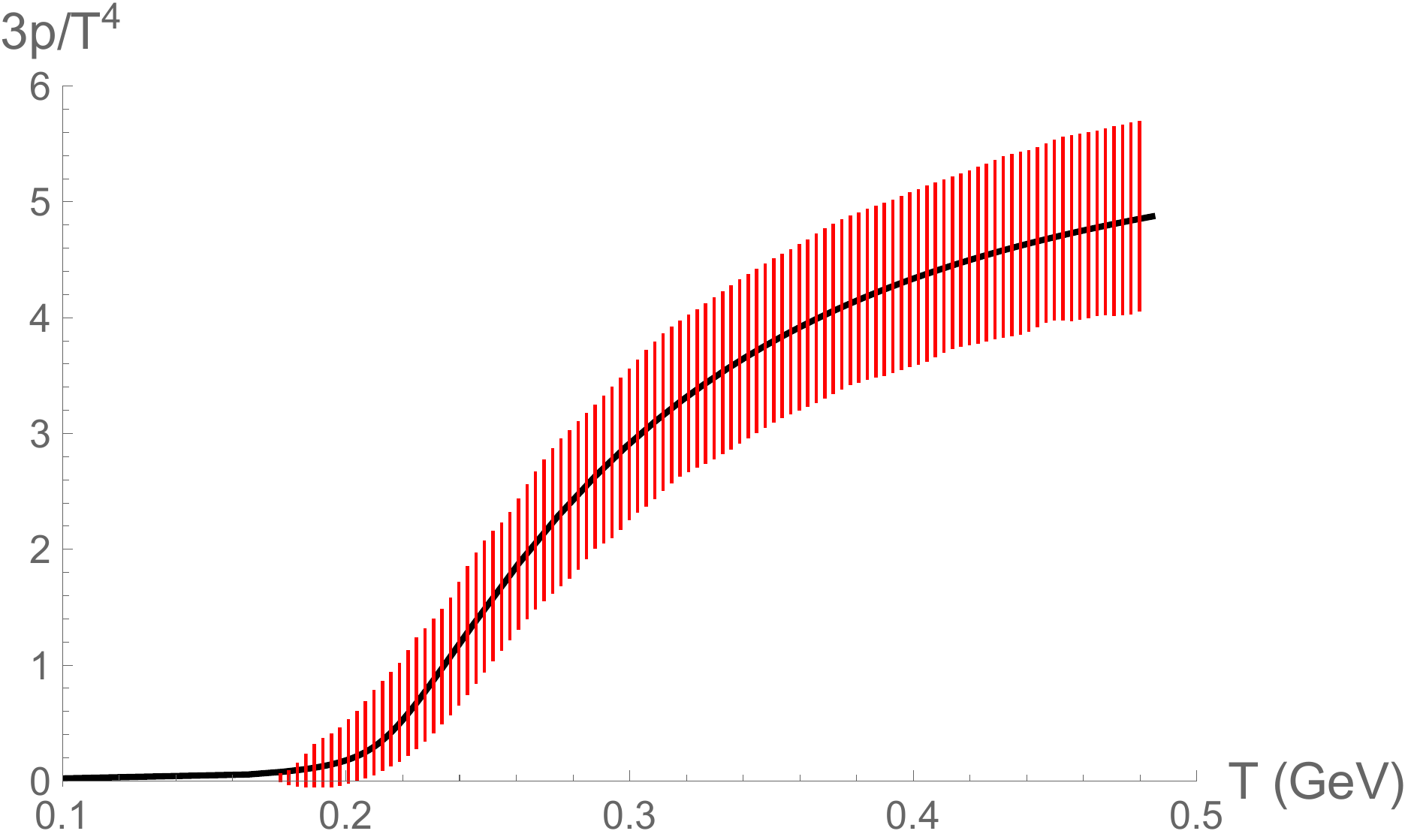}
\hspace*{0.6cm}
\includegraphics[width=68mm,clip=true,keepaspectratio=true]{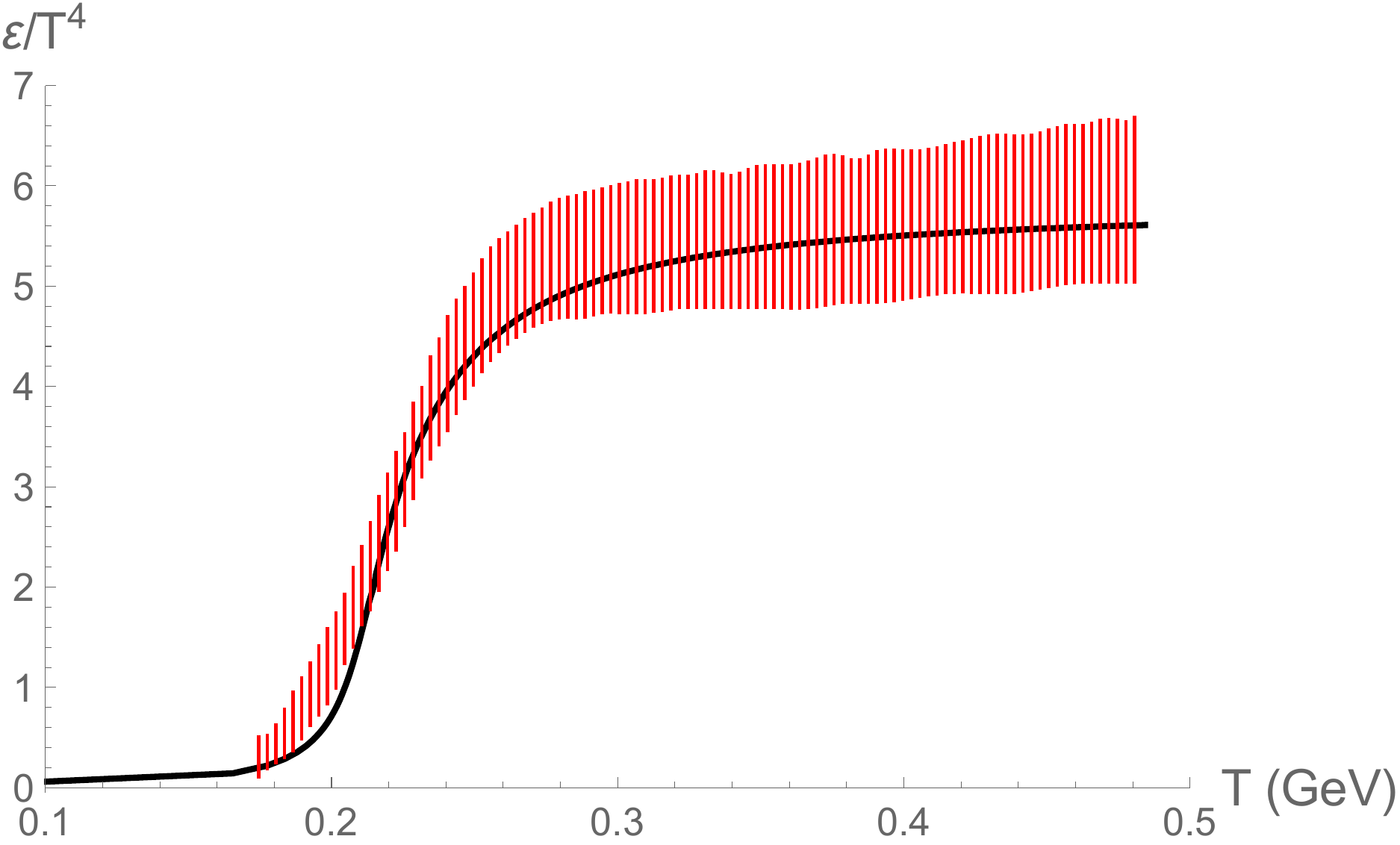}
\vskip -0.5cm \hskip 0.7 cm
\end{center}
\caption{The model results of the pressure $3p/T^4$ (left panel) and the energy density $\varepsilon/T^4$ (right panel) as functions of $T$ compared with the lattice interpolation results of two-flavor QCD denoted by the red band \cite{Burger:2014xga}.}
\label{p-e-T}
\end{figure*}

\begin{figure}
\centering
\includegraphics[width=75mm,clip=true,keepaspectratio=true]{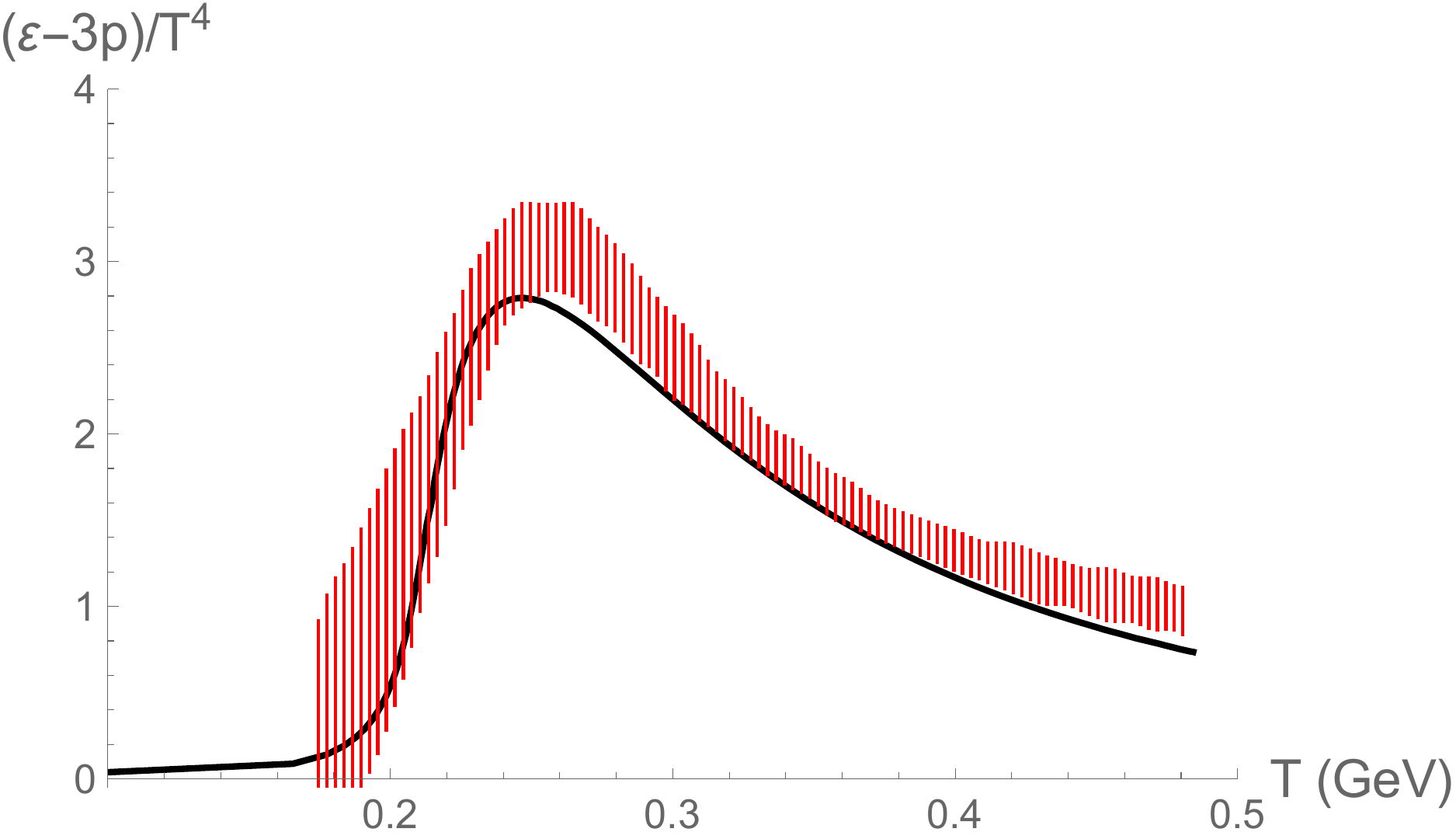}
\vskip 0.3cm
\caption{The model result of the trace anomaly $(\varepsilon-3p)/T^4$ as a function of $T$ compared with the lattice interpolation results of two-flavor QCD denoted by the red band \cite{Burger:2014xga}.} 
\label{e-3p-T}
\end{figure}

The temperature dependences of the entropy density $s/T^3$ and the speed of sound squared $c_s^2$ are shown in Fig. \ref{s-cs-T}, while in Fig. \ref{p-e-T} we compare the numerical results of the pressure $3p/T^4$ and the energy density $\varepsilon/T^4$ in unites of $T^4$ with the lattice interpolation results for the B-mass ensemble considered in Ref. \cite{Burger:2014xga}. In Fig. \ref{e-3p-T}, we present the model result of the trace anomaly $(\varepsilon-3p)/T^4$, which is also compared with the lattice interpolation result. We can see that the Einstein-dilaton system with a nontrivial dilaton potential can generate the crossover behavior of the equation of state which matches well with the lattice results.

\subsection{Polyakov loop}

The deconfining phase transition in thermal QCD is characterized by the VEV of the Polyakov loop which is defined as
\begin{align} \label{P-loop-def}
L(T) =\frac{1}{N_c} \tr P\exp\left[ig\int_0^{1/T}d\tau\hat{A}_0\right] ,
\end{align}
where $\hat{A}_0$ is the time component of the non-Abelian gauge field operator, the symbol $P$ denotes path ordering and the trace is over the fundamental representation of $SU(N_c)$. 

The VEV of the Polyakov loop in AdS/CFT is schematically given by the world-sheet path integral
\begin{align} \label{Polyakov-vev}
\VEV{L} =\int DX e^{-S_w},
\end{align}
where $X$ is a set of world-sheet fields and $S_w$ is the classical world-sheet action \cite{Andreev:2009zk,Noronha:2009ud}. In principle, $\VEV{L}$ can be evaluated approximately in terms of the minimal surface of the string world-sheet with given boundary conditions. In the low-energy and large $N_c$ limit, we have $\VEV{L}\sim e^{-S_{NG}}$ with the Nambu-Goto action
\begin{align} \label{S-NG}
S_{NG} =\frac{1}{2\pi\alpha'}\int d^2\sigma \sqrt{\det (g_{\mu\nu}^S\partial_a X^{\mu}\partial_b X^{\nu})} ,
\end{align}
where $\alpha'$ denotes the string tension, $g^S_{\mu\nu}$ is the string-frame metric and $X^{\mu}=X^{\mu}(\tau,\sigma)$ is the embedding of the world-sheet in the bulk spacetime. The regularized minimal world-sheet area takes the form
\begin{align} \label{reg-mini-area}
S_{R} =\frac{g_p}{\pi T} \int_\epsilon^{z_h}dz\frac{e^{2A_S}}{z^2}
\end{align}
with $g_p=\frac{L^2}{2\alpha'}$ \cite{Andreev:2009zk}. Subtracting the UV divergent terms and letting $\epsilon\to 0$, the renormalized world-sheet area can be obtained as
\begin{align}\label{ren-mini-area}
S_0 =S_0' +c_p =& \frac{g_p}{\pi T}\int_0^{z_h} dz \left[\frac{e^{2A_S}}{z^2} - \left(\frac{1}{z^2}+\frac{4p_1}{3z}\right)\right]       \nonumber\\
& +\frac{g_p}{\pi T}\left(\frac{4p_1}{3}\log z_h-\frac{1}{z_h}\right) +c_p ,
\end{align}
where $c_p$ is a scheme-dependent normalization constant.Thus the VEV of the Polyakov loop can be written as
\begin{align}\label{VEV-Ploop}
\VEV{L} = w e^{-S_0} = e^{-S_0' +c_p'},
\end{align}
where $w$ is a weight coefficient and the normalization constant $c_p'=\ln w-c_p$.

\begin{figure}
\centering
\includegraphics[width=75mm,clip=true,keepaspectratio=true]{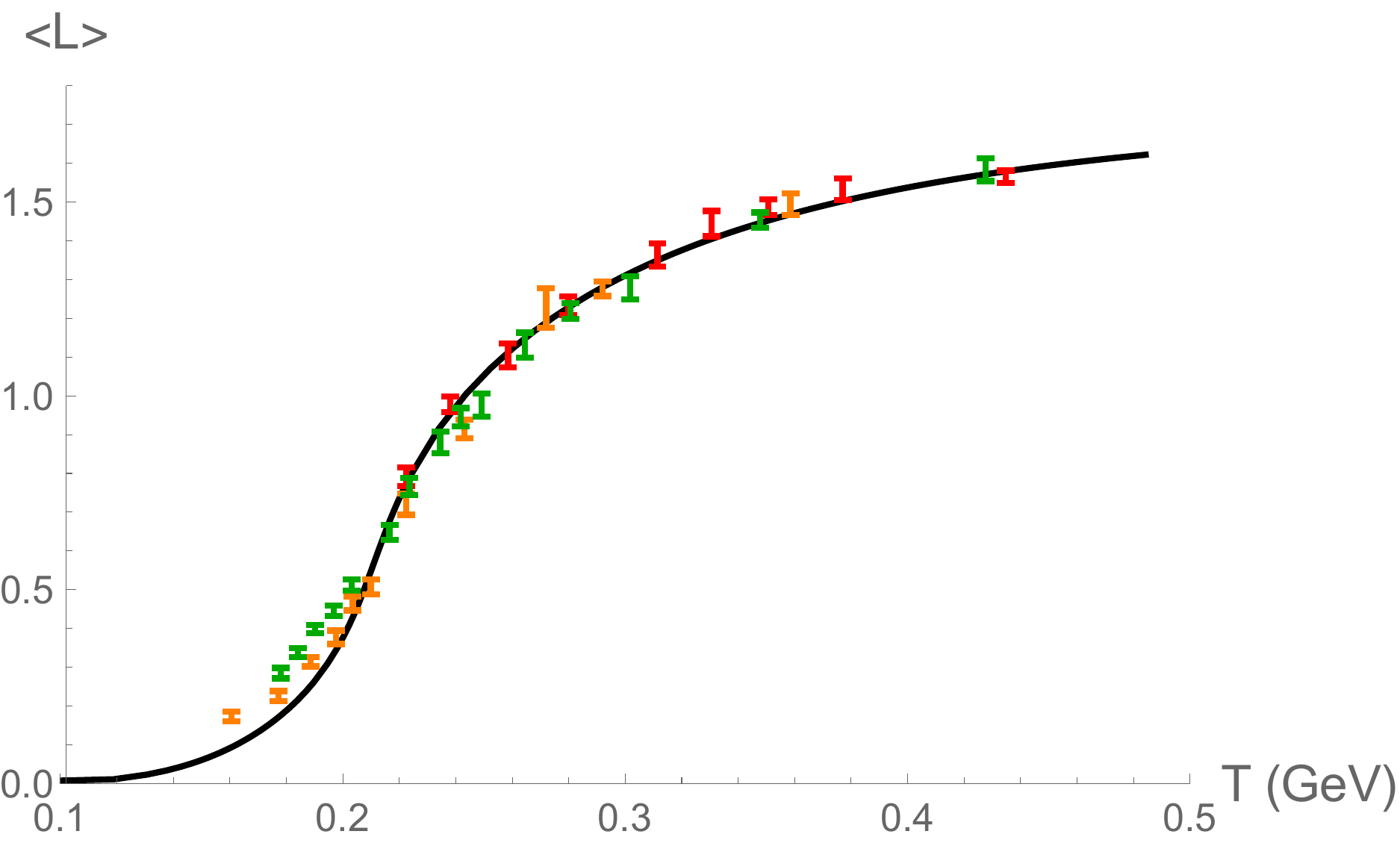}
\vskip 0.3cm
\caption{The model result of the VEV of Polyakov loop $\VEV{L}$ as a function of $T$ compared with the lattice data of the renormalized Polyakov loop for the B-mass ensemble denoted by the colored points with error bars \cite{Burger:2014xga}.} 
\label{L-T}
\end{figure}

We plot the temperature-dependent behavior of $\VEV{L}$ with the parameter values $g_p=0.29$ and $c_p'=0.16$ in Fig. \ref{L-T}, where we also show the two-flavor lattice data of the renormalized Polyakov loop (corresponding to the B-mass ensemble in \cite{Burger:2014xga}). We can see that the model result fits the lattice data quite well when we choose proper values of $g_p$ and $c_p'$.

\subsection{On the stability of the black hole solution}

One remark should be given on the background solution of the Einstein-dilaton system. In the above description of the equation of state, we have only used the black hole solution which is asymptotic to AdS$_5$ near the boundary, and have seen that this is crucial for the realization of the crossover transition. However, in principle, the Einstein-dilaton system also admits a thermal gas solution, which can be obtained by setting $f(z)=1$ \cite{Herzog:2006ra,Gursoy:2008za}. To guarantee the soundness of our calculation, we shall check the stability of the black hole solution against the thermal gas one.

According to AdS/CFT, the free energy is related to the on-shell action of the background fields by $\beta\mathcal{F} =\mathcal{S}^R$ with $\beta=1/T$, and the regularized on-shell action consists of three parts: 
\begin{align}\label{on-shell-act}
\mathcal{S}^R=\mathcal{S}_E+\mathcal{S}_{GH}+\mathcal{S}_{count} =\mathcal{S}_{\epsilon}+\mathcal{S}_{IR}+\mathcal{S}_{count},
\end{align}
where $\mathcal{S}_E$ denotes the on-shell Einstein-Hilbert action, $\mathcal{S}_{GH}$ denotes the Gibbons-Hawking term and $\mathcal{S}_{count}$ denotes the counter term. The subscripts $\epsilon$ and IR denote the contributions at UV cut-off $z=\epsilon$ and IR cut-off $z=z_{IR}$ respectively. Following Ref. \cite{Gursoy:2008za}, we can obtain the regularized on-shell action of the black hole (BH) solution:
\begin{align}\label{free-energy-BH1}
\mathcal{S}_{BH} =\mathcal{S}_{BH}^{\epsilon} +\mathcal{S}_{BH}^{count} =2\beta M^3V_3\left(3b^2(\epsilon)f(\epsilon)b'(\epsilon)+\frac{1}{2}b^3(\epsilon)f'(\epsilon)\right) +\mathcal{S}_{BH}^{count},
\end{align}
where $M^3\equiv1/(16\pi G_5)$, $V_3$ is the three-space volume and $b(z)\equiv\frac{L}{z}e^{A_E(z)}$. Note that $\mathcal{S}_{BH}$ has no IR contribution due to $f(z_h)=0$. The regularized on-shell action of the thermal gas (TG) solution takes the form:
\begin{align}\label{free-energy-TG1}
\mathcal{S}_{TG} &=\mathcal{S}_{TG}^{\epsilon} +\mathcal{S}_{TG}^{IR} +\mathcal{S}_{TG}^{count} =2\tilde{\beta}M^3\tilde{V}_3\left(3b_0^2(\epsilon)b_0'(\epsilon) +b_0^2(z_{IR})b_0'(z_{IR})\right) +\mathcal{S}_{TG}^{count},
\end{align}
where $b_0(z)\equiv\frac{L}{z}e^{A_{E0}(z)}$ and $\tilde{\beta}, \tilde{V}_3$ denote the corresponding quantities in the thermal gas case. To compare the free energies, we must make sure that the intrinsic geometries near the boundary should be the same for the two background solutions, i.e., the proper length of time circle and the proper volume of three-space should be the same at $z=\epsilon$, which imposes the following conditions \cite{Gursoy:2008za}:
\begin{align}\label{free-energ-condit}
\tilde{\beta}b_0(\epsilon)=\beta b(\epsilon)\sqrt{f(\epsilon)}, \quad \tilde{V_3}b_0^3(\epsilon)=V_3 b^3(\epsilon) .
\end{align}

With the condition (\ref{free-energ-condit}), the free energy difference between the two background solutions has the form:
\begin{align}\label{free-energy-dif1}
\mathcal{F} &=\beta^{-1}\lim_{\epsilon\to 0}(\mathcal{S}_{BH} -\mathcal{S}_{TG})    \nonumber \\
&=2M^3V_3\left(3b^2(\epsilon)f(\epsilon)b'(\epsilon)+\frac{1}{2}b^3(\epsilon)f'(\epsilon) -3\frac{b^4(\epsilon)}{b_0^2(\epsilon)}\sqrt{f(\epsilon)}b_0'(\epsilon)\right),
\end{align}
where the IR contribution in (\ref{free-energy-TG1}) has been omitted since this term vanishes for good singularities \cite{Gursoy:2008za}. In terms of the UV asymptotic forms (\ref{BG-uv-asymp}), we obtain the following result:
\begin{align}\label{free-energy-dif2}
\mathcal{F} =-\frac{1}{4}f_cL^3M^3V_3,
\end{align}
where we have taken the limit $\epsilon\to 0$. Note that the UV divergent terms in $\mathcal{S}_{BH}$ and $\mathcal{S}_{TG}$ have the same form, thus cancel in the final result. Since $f_c>0$, we have $\mathcal{F}<0$, which implies that the black hole phase is more stable than the thermal gas phase.

\section{Chiral transition in an improved soft-wall model with solved background}\label{sec-SW}

Our previous studies have shown that the chiral transition at zero baryon chemical potential can be characterized by an improved soft-wall AdS/QCD model in the AdS-Schwarzschild black hole background \cite{Fang:2016nfj,Fang:2018vkp}. However, this black hole solution cannot describe the QCD equation of state due to the conformal invariance of the dual gauge theory. Our main aim of this work is to combine the advantages of the improved soft-wall model in the description of chiral transition with a background system which can reproduce the deconfinement properties of QCD. As a first attempt, we investigate the possible ways to produce the chiral transition behaviors in the two-flavor case based on an improved soft-wall model (as the flavor part) under the more realistic background solved from the above Einstein-dilaton system.

\subsection{The flavor action}

We first outline the improved soft-wall AdS/QCD model with two flavors which is proposed in Ref. \cite{Fang:2016nfj}. The bulk action relevant to the chiral transition in this model is the scalar sector,
\begin{align}\label{flavor-action1}
S_X^{isw} &= -\int d^5x\sqrt{-g}e^{-\Phi}\mathrm{Tr}\left\{|\partial X|^2 +V_X(X)\right\} ,
\end{align}
where the dilaton takes the form $\Phi(z) =\mu_g^2\,z^2$ to produce the linear Regge spectra of light mesons, and the scalar potential is
\begin{align}\label{VX1}
V_X(X) =m_5^2(z)|X|^{2} +\lambda |X|^{4}
\end{align}
with a running bulk mass $m_5^2(z)=-3 -\mu_c^2\,z^2$. The constant term of $m_5^2(z)$ is determined by the mass-dimension relation $m_5^2L^2=\Delta(\Delta-4)$ for a bulk scalar field \cite{DaRold:2005mxj,Erlich:2005qh}, while the $z$-squared term is motivated by the phenomenology of meson spectrum and the quark mass anomalous dimension \cite{Fang:2016nfj}. 

In the holographic framework, a natural mechanism to produce such a $z$-dependent term of $m_5^2(z)$ is to introduce a coupling between the dilaton and the bulk scalar field. As we can see, without changing the results of the improved soft-wall model, the scalar potential can be recast into another form
\begin{align}\label{VX2}
V_X(X,\Phi) =m_5^2|X|^{2} -\lambda_1\Phi |X|^{2} +\lambda_2|X|^{4},
\end{align}
where $m_5^2=-3$, and a cubic coupling term of $\Phi$ and $X$ has been introduced. The effects of similar couplings on the low-energy hadron properties have also been considered in the previous studies \cite{Li:2013oda}. Here we propose such a change of $V_X$ from (\ref{VX1}) to (\ref{VX2}) with the aim to describe the chiral transition behaviors for the two-flavor case. Thus the flavor action that will be addressed in this work is
\begin{align}\label{flavor-action2}
S_X &= -\int d^5x\sqrt{-g}e^{-\Phi}\mathrm{Tr}\left\{|\partial X|^2 +V_X(X,\Phi)\right\} .
\end{align}

Unlike the previous studies, the metric and the dilaton in the flavor action (\ref{flavor-action2}) will be solved from the Einstein-dilaton system (\ref{AE-eom2}), (\ref{dilaton-eom2}) and (\ref{fz-eom3}), which has been shown to be able to reproduce the two-flavor lattice results of the equation of state. We shall assume that the flavor action (\ref{flavor-action2}) has been written in the string frame with the metric ansatz (\ref{stringmetric}). In our model, the dilaton field $\phi$ in the background action (\ref{gravity-act-str1}) has been distinguished from the field $\Phi$ in the flavor action (\ref{flavor-action2}). From AdS/CFT, these two fields may be reasonably identified as the same one, as indicated by the Dirac-Born-Infeld action which dictates the dynamics of the open string sector with the string coupling $g_s=e^{\phi}$. This is implemented in some works \cite{Li:2013oda}. However, the low-energy and large-$N$ limits taken in AdS/CFT and the further reduction to AdS/QCD has made things more subtle. The exact correspondence between $g_s$ and $e^{\phi}$ is not consolidated in the bottom-up AdS/QCD. On the other hand, the dilaton term $e^{-\Phi}$ in the flavor sector has been introduced to realize the Regge spectra of hadrons \cite{Karch:2006pv}. In this work, we concentrate on the phenomenological aspects and study how to realize more low-energy properties of QCD by the holographic approach. Thus we try a more general form $\Phi=k\phi$ with $k$ a parameter, which will not affect the linear Regge spectra qualitatively. In the actual calculation, we just choose two simplest cases $k=1$ and $k=2$ to investigate the effects of $k$ on chiral behaviors. The probe approximation which neglects the backreaction effect of the flavor sector on the background system will be adopted in this work, as in the most studies on AdS/QCD with fixed background.

\subsection{The EOM of the scalar VEV}

According to AdS/CFT, the VEV of the bulk scalar field in the two-flavor case can be written as $\langle X\rangle=\frac{\chi(z)}{2}I_2$ with $I_2$ denoting the $2\times2$ identity matrix, and the chiral condensate is incorporated in the UV expansion of the scalar VEV $\chi(z)$ \cite{Erlich:2005qh}. To address the issue on chiral transition, we only need to consider the action of the scalar VEV,
\begin{align}\label{chi-act-str1}
S_{\chi} &= -\int d^5x\sqrt{-g}e^{-\Phi}\left(\frac{1}{2}g^{zz}(\partial_z\chi)^2 +V(\chi,\Phi)\right)
\end{align}
with
\begin{align}\label{Vchi1}
V(\chi,\Phi) &=\mathrm{Tr}\,V_X(\VEV{X},\Phi) =\frac{1}{2}(m_5^2-\lambda_1\Phi)\chi^{2} +\frac{\lambda_2}{8} \chi^{4} .
\end{align}

In terms of the metric ansatz (\ref{stringmetric}), the EOM of $\chi(z)$ can be derived from the action (\ref{chi-act-str1}) as
\begin{align}\label{str-chi-eom1}
\chi''+\left(3A_S'-\Phi' +\frac{f'}{f}-\frac{3}{z}\right)\chi' -\frac{e^{2A_S}\,\partial_\chi V(\chi,\Phi)}{z^2 f} =0 .
\end{align}
The UV asymptotic form of $\chi(z)$ at $z\to0$ can be obtained from Eq. (\ref{str-chi-eom1}) as
\begin{align}\label{chi-uv}
\chi(z) =\,& m_q\zeta z +(6-k+k\lambda_1)m_q p_1\zeta z^2 +\frac{\sigma}{\zeta}z^3        \nonumber\\ 
&+\frac{1}{3}\left[m_q\zeta p_1^2\left(30 k -3k^2 -23k\lambda_1 -\frac{3}{2}k^2\lambda_1^2 \right.\right.       \nonumber\\
&\left.\left. +\frac{9}{2}k^2\lambda_1 -\frac{224}{3} \right) +\frac{3}{4}m_q^3\zeta^3\lambda_2\right] z^3\log z +\cdots ,
\end{align}
where $m_q$ is the current quark mass, $\sigma$ is the chiral condensate, and $\zeta=\frac{\sqrt{N_c}}{2\pi}$ is a normalization constant \cite{Cherman:2008eh}. As in Eq. (\ref{dilaton-eom2}), a natural boundary condition at horizon $z_h$ follows from the regular condition of $\chi(z)$ near $z_h$,
\begin{align} \label{chi-eom-BC}
\left[f'\chi' -\frac{e^{2A_S}}{z^2}\partial_\chi V(\chi,\Phi)\right]_{z=z_h} =0 .
\end{align}

\subsection{Chiral transition}

To study the chiral transition properties in the improved soft-wall AdS/QCD model with the given background, we need to solve the scalar VEV $\chi(z)$ numerically from Eq. (\ref{str-chi-eom1}) with the UV asymptotic form (\ref{chi-uv}) and the IR boundary condition (\ref{chi-eom-BC}). The chiral condensate can then be extracted from the UV expansion of $\chi(z)$. In the calculation, we will take the set of parameter values which has been used to fit the lattice results of the equation of state in the two-flavor case (see Sec. \ref{sec-EOS}), and the quark mass will be fixed as $m_q=5\MeV$.

\begin{figure}
\centering
\includegraphics[width=75mm,clip=true,keepaspectratio=true]{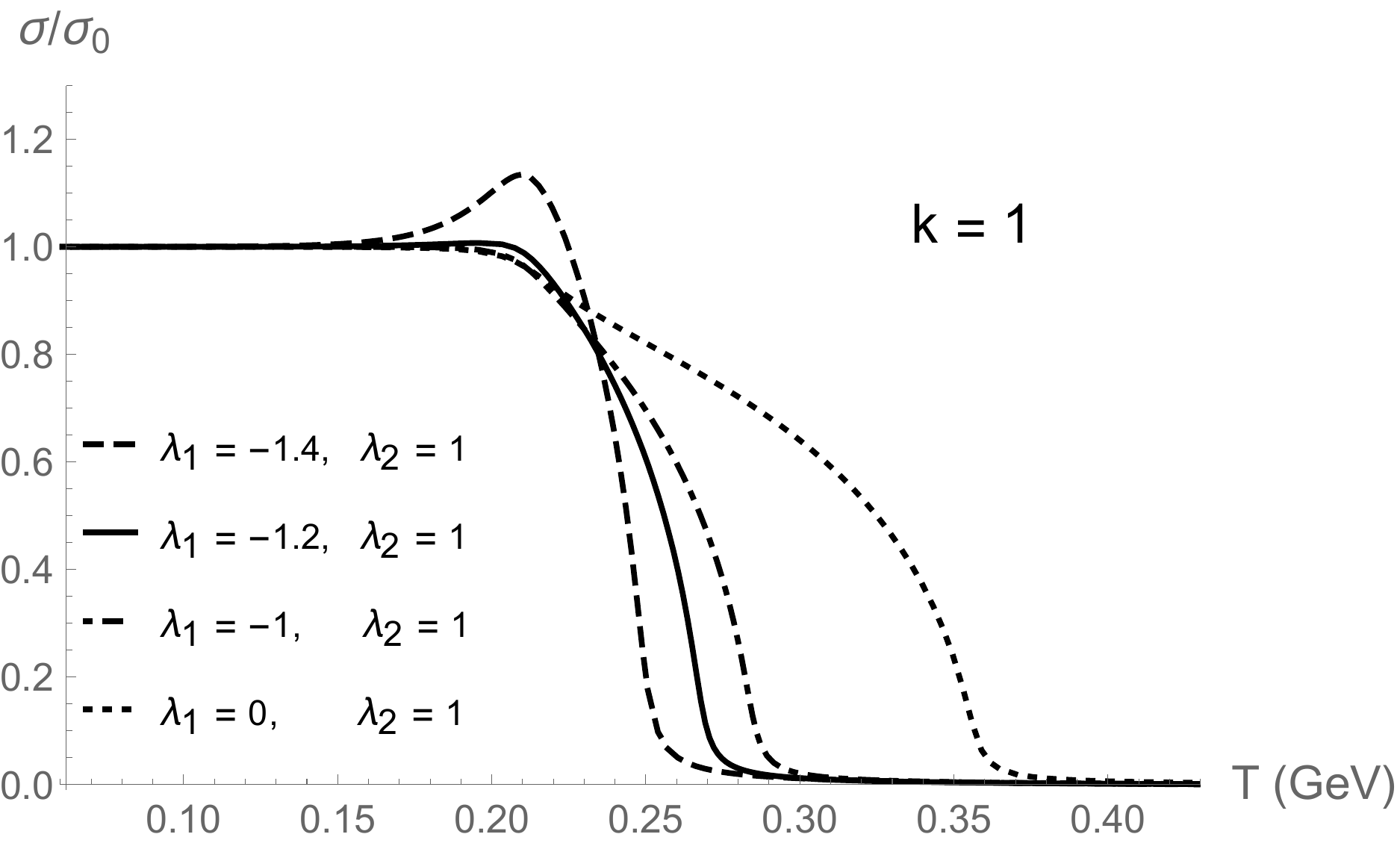}
\hspace*{0.6cm}
\includegraphics[width=75mm,clip=true,keepaspectratio=true]{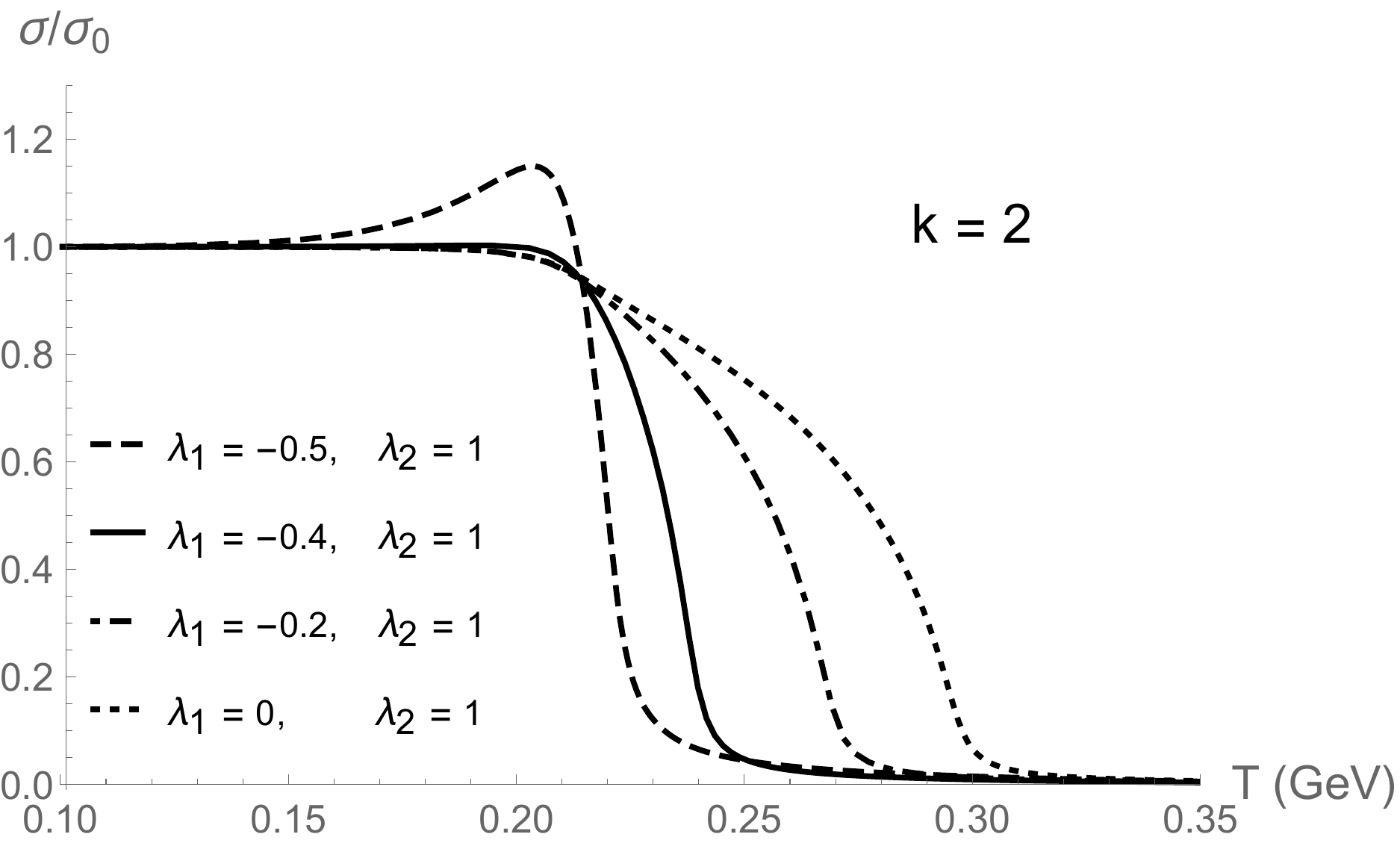}
\vskip 0.3cm
\caption{The normalized chiral condensate $\sigma/\sigma_0$ as a function of $T$ for different values of $\lambda_1$ in the cases $k=1$ and $k=2$ with $\lambda_2=1$.} 
\label{fig-sigma-T1}
\end{figure}

In this work, we only consider two cases corresponding to $k=1$ ($\Phi=\phi$) and $k=2$ ($\Phi=2\phi$). In each case, the temperature dependence of the chiral condensate normalized by $\sigma_0=\sigma(T=0)$ will be investigated for a set of values of $\lambda_1$ and $\lambda_2$. We first fix $\lambda_2=1$, and select four different values of $\lambda_1$ for each case. The model results of the normalized chiral condensate $\sigma/\sigma_0$ as a function of $T$ are shown in Fig. \ref{fig-sigma-T1}. We can see that the crossover transition can be realized qualitatively in such an improved soft-wall model with the solved gravitational background. Moreover, we find that there is a decreasing tendency for the transition temperature with the decrease of $\lambda_1$, yet a visible bump emerges near the transition region at relatively smaller $\lambda_1$ and only disappears gradually with the increase of $\lambda_1$. As shown in Fig. \ref{fig-sigma-T1}, we find that the transition temperatures with our selected parameter values are larger than the lattice result $T_{\chi}\sim193\MeV$ \cite{Burger:2014xga}.

\begin{figure}
\centering
\includegraphics[width=75mm,clip=true,keepaspectratio=true]{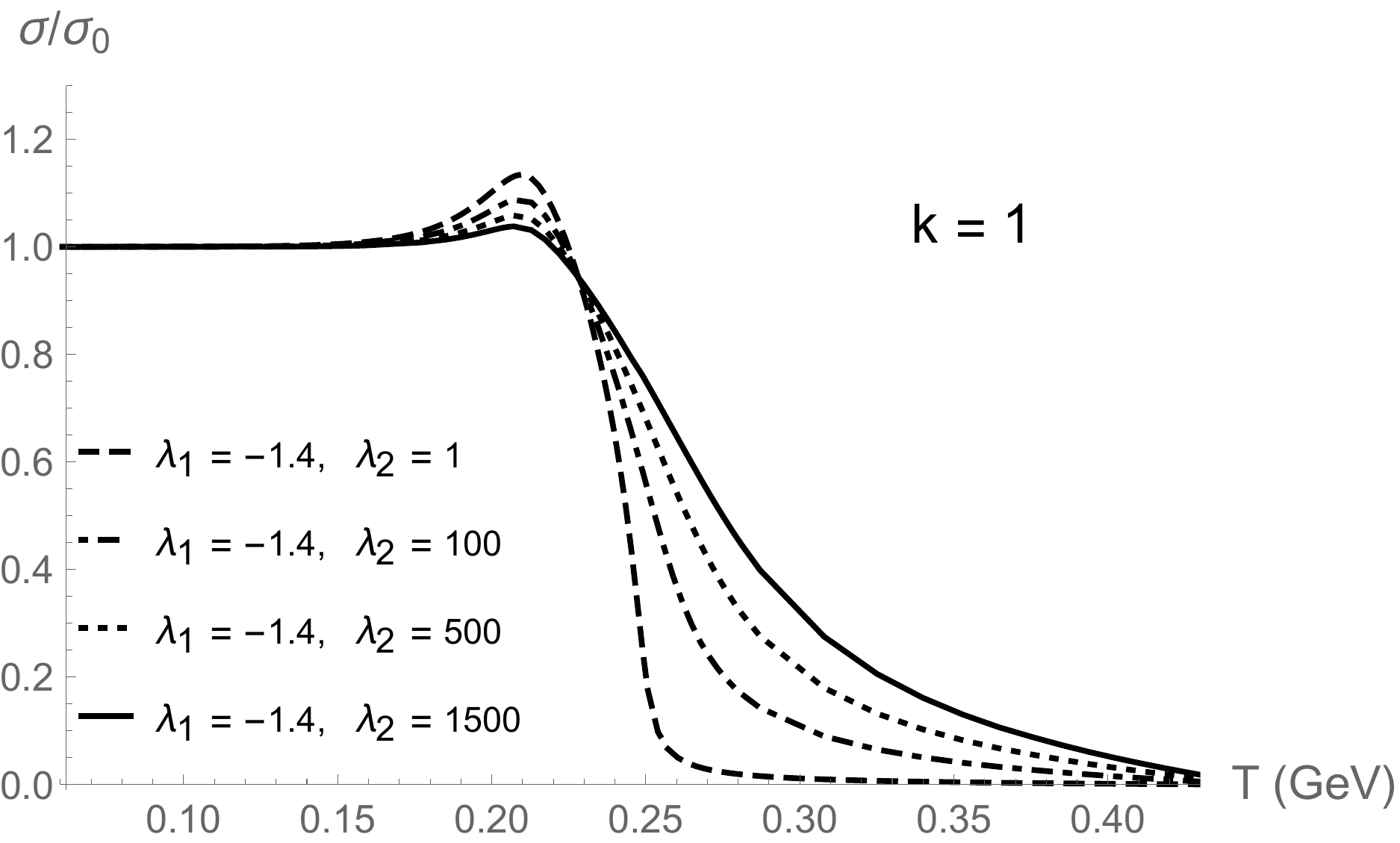}
\hspace*{0.6cm}
\includegraphics[width=75mm,clip=true,keepaspectratio=true]{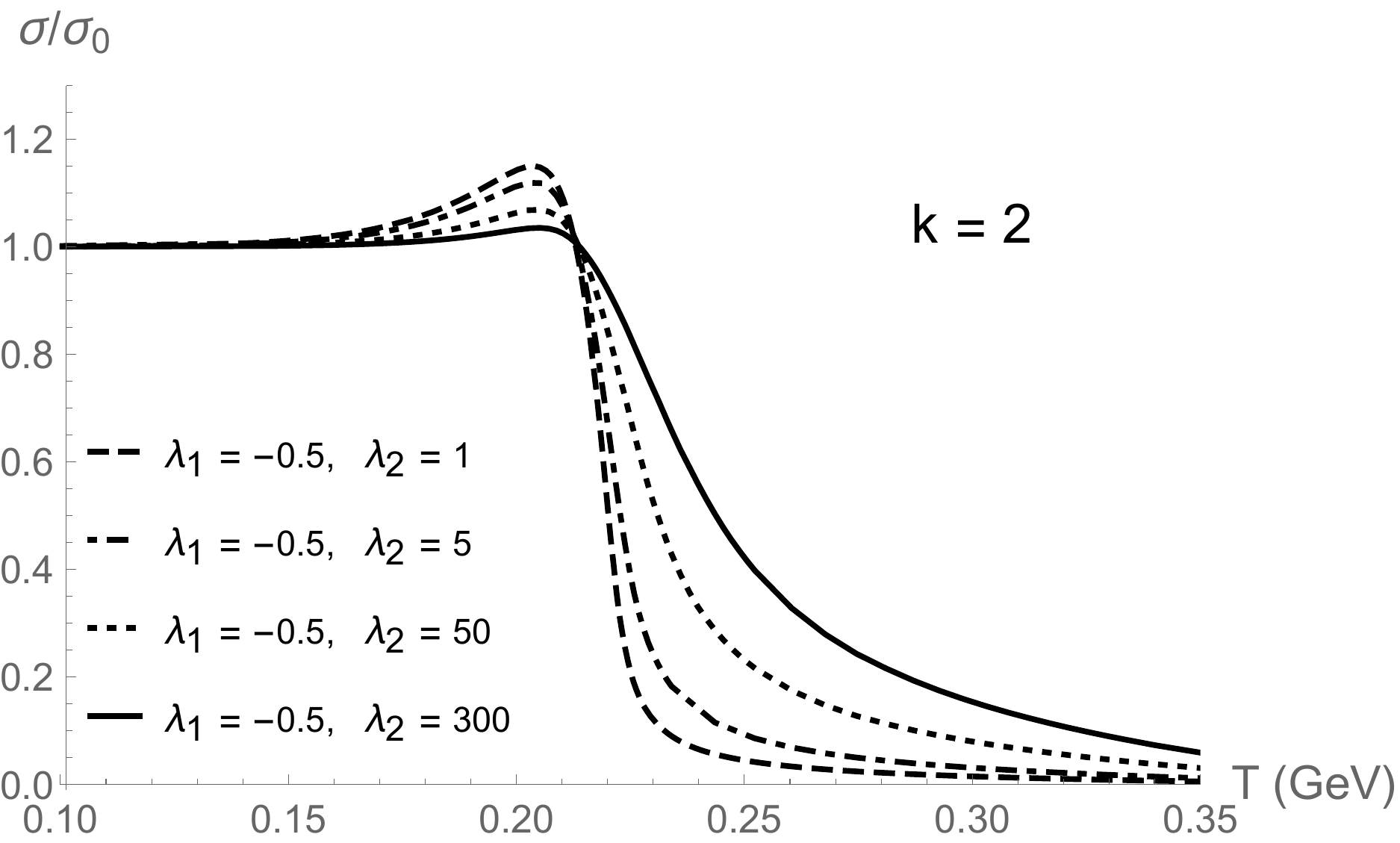}
\vskip 0.3cm
\caption{The normalized chiral condensate $\sigma/\sigma_0$ as a function of $T$ for different values of $\lambda_2$ in the cases $k=1$ and $k=2$ with fixed values of $\lambda_1$.} 
\label{fig-sigma-T2}
\end{figure}

We then investigate the effects of the quartic coupling constant $\lambda_2$ on chiral transitions. We fix $\lambda_1=-1.4$ for the case $k=1$ and $\lambda_1=-0.5$ for the case $k=2$, and select four different values of $\lambda_2$ in each case. The chiral transition curves are plotted in Fig. \ref{fig-sigma-T2}. The result shows that with the increase of $\lambda_2$ the bump near the transition region becomes smaller and the normalized chiral condensate $\sigma/\sigma_0$ descends gently with $T$, though the value of $\lambda_2$ needs to be very large in order to smooth out the bump.

\section{Conclusion and discussion}\label{sec-conc}

We considered an improved soft-wall AdS/QCD model with a cubic coupling term between the dilaton and the bulk scalar field in a more realistic background, which is solved from the Einstein-dilaton system with a nontrivial dilaton potential. Such an Einstein-dilaton system has been used to reproduce the equation of state from lattice QCD with two flavors. Then the chiral transition behaviors were investigated in the improved soft-wall model based on the solved bulk background. We only considered two typical cases with $k=1$ and $k=2$, and the quartic coupling constant is firstly fixed as $\lambda_2=1$. In both cases, the crossover behavior of chiral transition can be realized, as seen from Fig. \ref{fig-sigma-T1}. Nevertheless, the chiral transition temperatures obtained from the model are much larger than the lattice result. Although $T_{\chi}$ decreases with the decrease of $\lambda_1$, a visible bump near the transition region will emerge when $\lambda_1$ is small enough. We then studied the influence of the value of $\lambda_2$ on chiral transition, as shown in Fig. \ref{fig-sigma-T2}. We find that in some sense the quartic coupling term can smooth the bump, but to remove it, the value of $\lambda_2$ needs to be very large.

In our consideration, the scaling dimension of the dual operator $\tr F^2_{\mu\nu}$ of the dilaton has been taken as $\Delta=3$, which can be used to mimick the QCD equation of state \cite{Finazzo:2013efa,Finazzo:2014zga}. However, we remark that the properties of thermal QCD considered in our work are not determined exclusively by one particular value of $\Delta$. Indeed, other values of $\Delta$ have also been adopted for the realization of equation of state with slightly different forms of $V_E(\phi)$ \cite{Gubser:2008yx,Noronha:2009ud}. Since the UV matching to QCD at a finite scale cannot capture asymptotic freedom, we are content to give a phenomenological description on the thermodynamic properties of QCD, which are expected not so sensitive to the UV regime from the angle of renormalization and effective field theory. In this work, we have built an improved soft-wall AdS/QCD model under a more realistic gravitational background, which provides a possibility in the holographic framework to address the deconfining and chiral phase transition simultaneously.

We have assumed that the backreaction of the flavor sector to the background is small enough such that we can just adopt the solution of the Einstein-dilaton system as the bulk background under which the chiral properties of the improved soft-wall model are considered. This is sensible only when we take a small weight of the flavor action (\ref{flavor-action2}) compared to the background action (\ref{gravity-act-str1}). To clarify the phase structure in this improved soft-wall AdS/QCD model, we shall consider the backreaction of the flavor part to the background system thoroughly. The correlation between the deconfining and chiral phase transitions can then be studied in such an improved soft-wall model coupled with an Einstein-dilaton system. The case of finite chemical potential can also be considered by introducing a $U(1)$ gauge field in order to study the properties of QCD phase diagram.

\section*{Acknowledgements}
This work is supported in part by the National Natural Science Foundation of China (NSFC) under Grant Nos. 11851302, 11851303, 11690022 and 11747601, and the Strategic Priority Research Program of the Chinese Academy of Sciences under Grant No. XDB23030100 as well as the CAS Center for Excellence in Particle Physics (CCEPP). Z. F. is also supported by the NSFC under Grant No. 11905055 and the Fundamental Research Funds for the Central Universities under Grant No. 531118010198.

\bibliography{refs-AdSQCD}

\end{document}